\documentclass[12pt,preprint,a4paper]{aastex}
\usepackage{graphics,epsf}
\usepackage{amsmath}                
\usepackage{amsfonts}               
\usepackage{amssymb}                
\usepackage{epsfig}                 
\usepackage{subfigure}
\usepackage{float}

\def \cm{~\rm{cm}}
\def \s{~\rm{s}}
\def \km{~\rm{km}}

\def \K{~\rm{K}}
\def \g{~\rm{g}}

\def \AU{~\rm{AU}}
\def \erg{~\rm{erg}}

\begin{document}

\title{ACCELERATING VERY FAST GAS IN THE SUPERNOVA IMPOSTOR SN~2009ip WITH JETS FROM A STELLAR COMPANION}

\author{Danny Tsebrenko\altaffilmark{1} and Noam Soker\altaffilmark{1}}

\altaffiltext{1}{Department of Physics, Technion -- Israel Institute of Technology, Haifa 32000, Israel;
ddtt@tx.technion.ac.il; soker@physics.technion.ac.il.}
\begin{abstract}
Using hydrodynamical numerical simulations we show that high-velocity ejecta with $v \sim 10^4 \km \s^{-1}$
in the outbursts of the supernova impostor SN~2009ip and similar luminous blue variable (LBV) stars can be explained
by the interaction of fast jets, having $v_{\rm jet} \sim 2000-3000 \km \s^{-1}$, with a circumbinary shell (extended envelope).
The density profile in the shell is very steep such that the shock wave, that is excited by the jets' interaction with the shell, accelerates
to high velocities as it propagates outward.
The amount of very fast ejecta is small, but sufficient to account for some absorption lines.
Such an extended envelope can be formed from the binary interaction and/or the unstable phase of the LBV primary star.
The jets themselves are launched by the more compact secondary star near periastron passages.
\end{abstract}


\section{INTRODUCTION}
\label{sec:introduction}
The SN impostor SN~2009ip is an LBV star \citep{Berger2009} that experienced a series
of outbursts starting in 2009 
(e.g., \citealt{Maza2009,Drake2012, Levesque2012, Mauerhan2013, Pastorello2012}).
There were two outbursts in 2012: the 2012a outburst that was similar to the previous outbursts,
and the 2012b outburst in September 2012 that was
much more energetic than the previous outbursts (e.g., \citealt{Mauerhan2013}).
There is no consensus on the nature of the 2012a and 2012b outbursts
(e.g. \citealt{Martinetal2013, Smithetal2013, Ouyedetal2013}), and not even on whether the LBV primary star
survived the 2012 outbursts or exploded as a core collapse supernova (CCSN).

The peak luminosity of the 2012b outburst and the P-Cygni absorption wings that extend
to $\sim 13\,000 \km \s^{-1}$ brought \cite{Mauerhan2013}
to suggest that the 2012a outburst was a weak SN event,
while the major 2012b outburst is a result of the collision of the SN ejecta with previously ejected gas (also \citealt{Prieto2013}).
However, \cite{Marguttietal2013} noted that the photosphere expansion velocity of $\sim 4500 \km \s^{-1}$
during the 2012b outburst implies that the gas that accelerated the photosphere must have originated long after the peak of the 2012a event.
Namely, the gas was ejected long after the star has ceased to exist according to the scenario proposed by \cite{Mauerhan2013}.
\cite{Martinetal2013} suggest that the star did survive the 2012 outbursts due to the fact that
peaks and troughs are present in the declining light curve of the 2012b outburst.

\cite{Pastorello2012} noted that the September 2011 outburst already had such wide P-Cygni wings,
and raised the possibility that the 2012b outburst was not a SN after all.
The early outbursts of SN~2009ip were compared with the 1837--1856 Great Eruption of $\eta$ Carinae
\citep{Smithetal2010, Smithetal2011, Foleyetal2011, Mauerhan2013, Levesque2012, SokerKashi2013}.
$\eta$ Carinae also had a small fraction of the mass ejected during the Great Eruption moving at high velocities
 \citep{Weisetal2001} of up to $\sim 5000 \km \s^{-1}$ \citep{Smith2008}.
\cite{SokerKashi2013}, for a terminal periastron passage,  and \cite{Kashietal2013}, for repeated periastron passages,
went further and argued that the 2012b outburst was powered by mass accretion onto a main sequence (MS)
or a Wolf-Rayet (WR) stellar companion, much like their model for powering the Great Eruption of $\eta$ Carinae \citep{KashiSoker2010}.
In the scenarios considered by \cite{Kashietal2013} the major 2012b outburst of $\sim 5 \times 10^{49} \erg$
was powered by $\sim 5 M_\odot$ accreted onto the companion during a periastron passage
(the first passage) of the binary system when the companion was inside the extended primary envelope.

The major open question in the binary scenarios for SN~2009ip is how such an interaction can give rise
to ejecta with velocities of $\ga 10\,000 \km \s^{-1}$.
In this first study we limit ourselves to outbursts with low-mass high-velocity ejecta,
such as the 2011 outburst of SN 2009ip and the Great Eruption of $\eta$ Carinae.
The numerical setup is described in section \ref{sec:numerics}.
Our results are described in section \ref{sec:results}, and our summary is in section \ref{sec:summary}.

\section{NUMERICAL SETUP}
\label{sec:numerics}

\begin{figure}[ht!]
\begin{center}
\includegraphics*[scale=0.5,clip=false,trim=0 0 0 0]{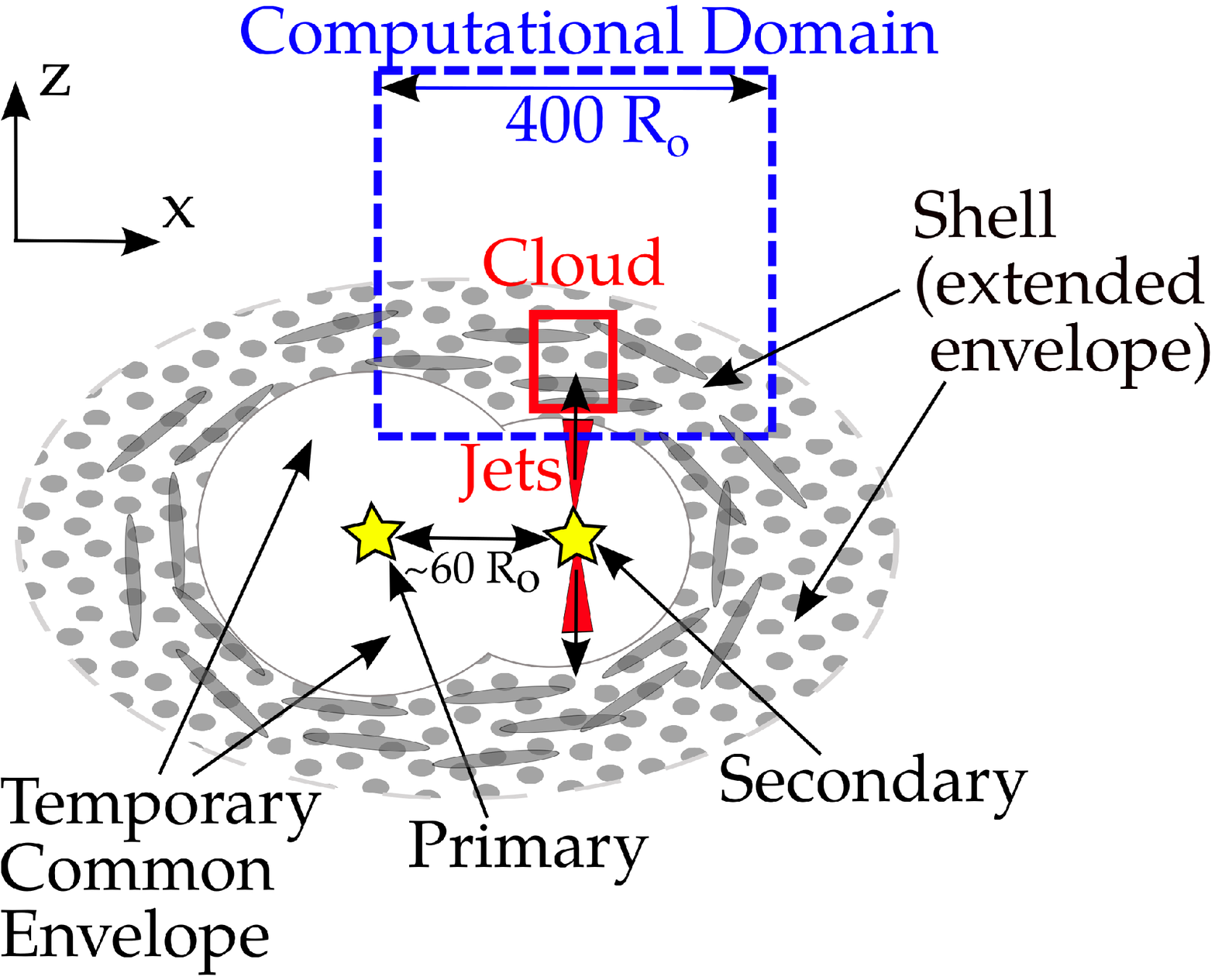}\\
\includegraphics*[scale=0.15,clip=false,trim=0 0 0 0]{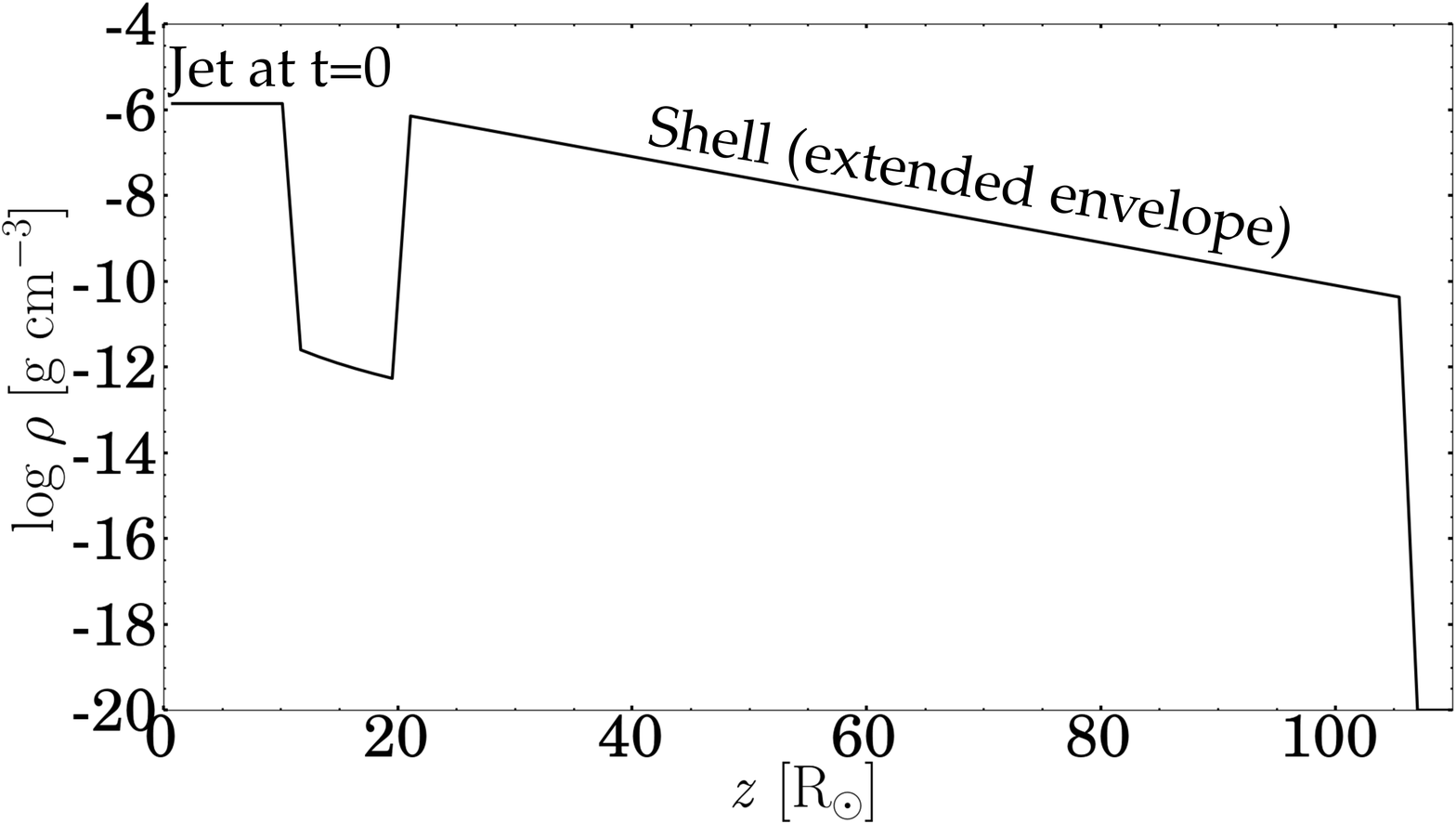}
\caption{
Upper panel: A schematic drawing of the initial flow setup in the $y = 0$ plane.
The blue dashed line marks the computational domain.
The solid red line marks the initial location of the cloud with which the jet interacts.
Lower panel: The density profile along the $z$ axis through the initial jet launching cells and the shell at $t = 0$
(before jets' expansion took place) for the $v_{\rm{jet}} = 3000 \km \s^{-1}$ run.
The simulation results for this density profile are presented in Figures \ref{fig:results_3000} and \ref{fig:vel_hist}.
Note that $z=0$ is not the equatorial plane, but rather the lower boundary of the computational domain. }
\label{fig:setup}
\end{center}
\end{figure}

The simulations are performed using the hydrodynamics code {\sc{flash 4.0}} \citep{Fryxell2000}.
We employ a full 3D uniform grid (equal-size cells) with Cartesian $(x,y,z)$ geometry.
In all runs the grid is a cube of size $400 R_\odot$ composed of $256^3$ cubical cells.
A high-resolution run with $512^3$ cubical cells shows no difference in the results. 
The temporary common envelope has a general oblate structure \citep{RickerTaam2012, Passyetal2012},
that allows us to use a rectangular grid and a simple plane-parallel structure for the shell,
schematically presented in Figure \ref{fig:setup}.

We simulate the general flow structure proposed by \cite{Kashietal2013},
where the companion launches two opposite collimated outflows while inside or close to the primary star.
The $z = 0$ plane is the lower boundary of the computational domain, just above the outskirts of the temporary common envelope,
${\rm several} \times 10 R_\odot$ above the equatorial plane.
The symmetry of the flow allows us to simulate the jet-shell interaction only in the $z>0$  half space.
The jet is injected along the $+z$ direction and from the bottom three cells of the computational grid.

For the 2012b outburst of SN~2009ip \cite{Kashietal2013} suggest that the interaction lasted for $\sim 10$~days during which the companion accreted
$\sim 5 M_\odot$.
We consider the less energetic 2011 outburst and simulate only $\sim 6-12$~hours of jets' launching episode, but assume a similar accretion rate.
We also take most of the gravitational energy released by the accretion process to be channelled to the jets,
such that the kinetic power in each jet is $P_{\rm jet} \simeq 6 \times 10^{43} \erg \s^{-1} $.
The jets velocity is taken to be $v_{\rm{jet}} = 3000 \km \s^{-1}$, similar to the wind velocity of the secondary in $\eta$ Carinae \citep{PittardCorcoran2002}.
We also simulate a case with the same power but with $v_{\rm{jet}} = 2000 \km \s^{-1}$,
similar to the escape velocity from very massive MS and WR stars.
The total ejected mass in the two jets in our simulation is $\sim 0.02 M_\odot$ ($\sim 0.06 M_\odot$)
for the jets velocity of $v_{\rm{jet}}=3000 \km \s^{-1}$ at 8 hours simulation time ($v_{\rm{jet}}=2000 \km \s^{-1}$ at 11 hours simulation time).
The secondary is a MS star, and we take the jets radius to be $10 R_\odot$, $\sim 1.5$ times the secondary radius.
As the companion accretes only $\sim 0.2-0.3 M_\odot$, and by our assumption most of the released gravitational energy is carried by the jets, we don't expect the radius of the secondary star to change by much for the duration of the simulation.

We assume that the jets interact with a shell, that might actually be an extended envelope.
Such an extended envelope might be formed around unstable LBV and AGB stars experiencing high mass loss rates (e.g., \citealt{ Soker2008}).
The LBV star $\eta$ Carinae had a huge atmosphere during its Great Eruption, when the photosphere was at $\sim 10 \AU$ \citep{DavidsonHumphreys1997}.
The radius of pulsating AGB stars can vary by a factor of $\sim 2-3$ (e.g., \citealt{Ireland2011, Ohnaka2013}).
Such an extended envelope is not a wind, and it is expected to be rather smooth and not clumpy as winds might be.
In the proposed model the extended envelope results both from the unstable phase
of the LBV primary star and from the tidal disturbance caused by the companion.
We assume that this expansion of the envelope results in a steep density profile.
One might take the steep density profile as a requirement of our model to obtain high velocities. 

The density profile of the shell is taken to be
\begin{equation}
\rho_{\rm{shell}}(z) = \rho_0 \times 10^{- \left[ ({z - z_{\rm{i}}})/{z_{\rm{scale}}} \right]} \qquad {\rm for} \qquad 20 R_{\odot} < z < 100 R_{\odot},
\label{eq:density}
\end{equation}
where $z_{\rm{scale}} =20 R_{\odot}$, $z_{\rm{i}} = 20 R_{\odot}$ is the location of the lower boundary of the shell and
$\rho_0 = 8 \times 10^{-7} \g \cm^{-3}$ ($\rho_0 = 18 \times 10^{-7} \g \cm^{-3}$) for a jet velocity of
$v_{\rm{jet}}=3000 \km \s^{-1}$ ($v_{\rm{jet}}=2000 \km \s^{-1}$).
The initial temperature of the shell is set to $T = 10^4 \K$.
We let the jet interact with a `cloud' (see Figure \ref{fig:setup}), namely, part of the shell.
When we took a full shell (extended in the entire $x-y$ directions) we obtained identical results.

For numerical stability we allow a low density gas,
$\rho_{\rm gas} = 10^{-10} (z/2R_\odot)^{-3} \g \cm^{-3}$,  to fill the space between the initial jet and the shell (cloud) at $t=0$,
extended in the entire $x - y$ plane between $ z = 0$ and $ z = 30 R_{\odot} $.

The ambient density outside the shell is set to be $ \rho_{\rm{amb}} = 10^{-20} \g \cm^{-3}$ at a temperature of $10^4 \K$.
As the material in the shell heats up to high temperatures, $T \ga 10^8\K$,
radiation pressure dominates and we take the adiabatic index to be $\gamma = 4/3$.
\section{RESULTS}
\label{sec:results}
\begin{figure}[ht!]
\begin{center}
\subfigure{\label{subfigure:dens1}\includegraphics*[scale=0.21,clip=true,trim=0 0 0 0]{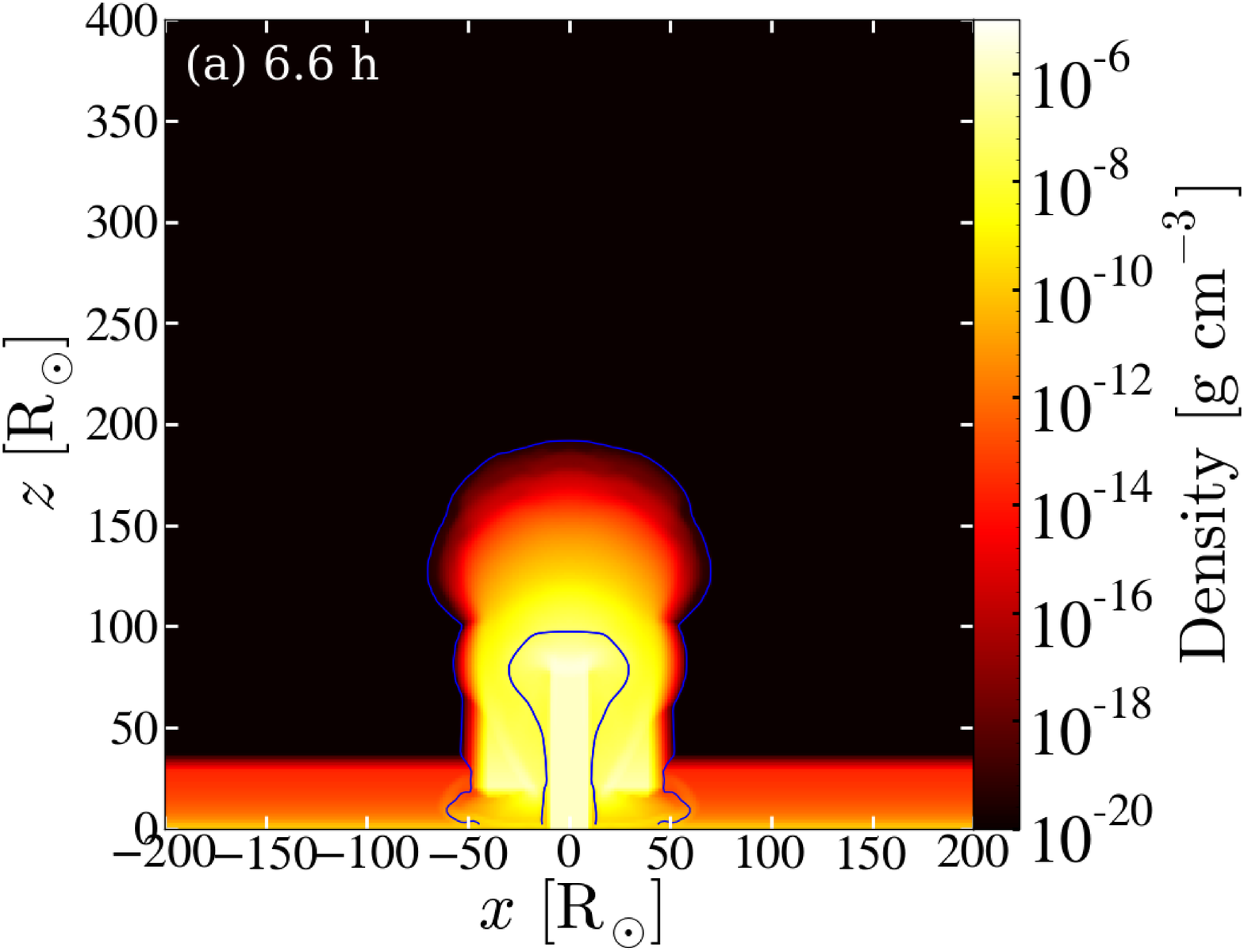}}
\subfigure{\label{subfigure:dens2}\includegraphics*[scale=0.21,clip=true,trim=0 0 0 0]{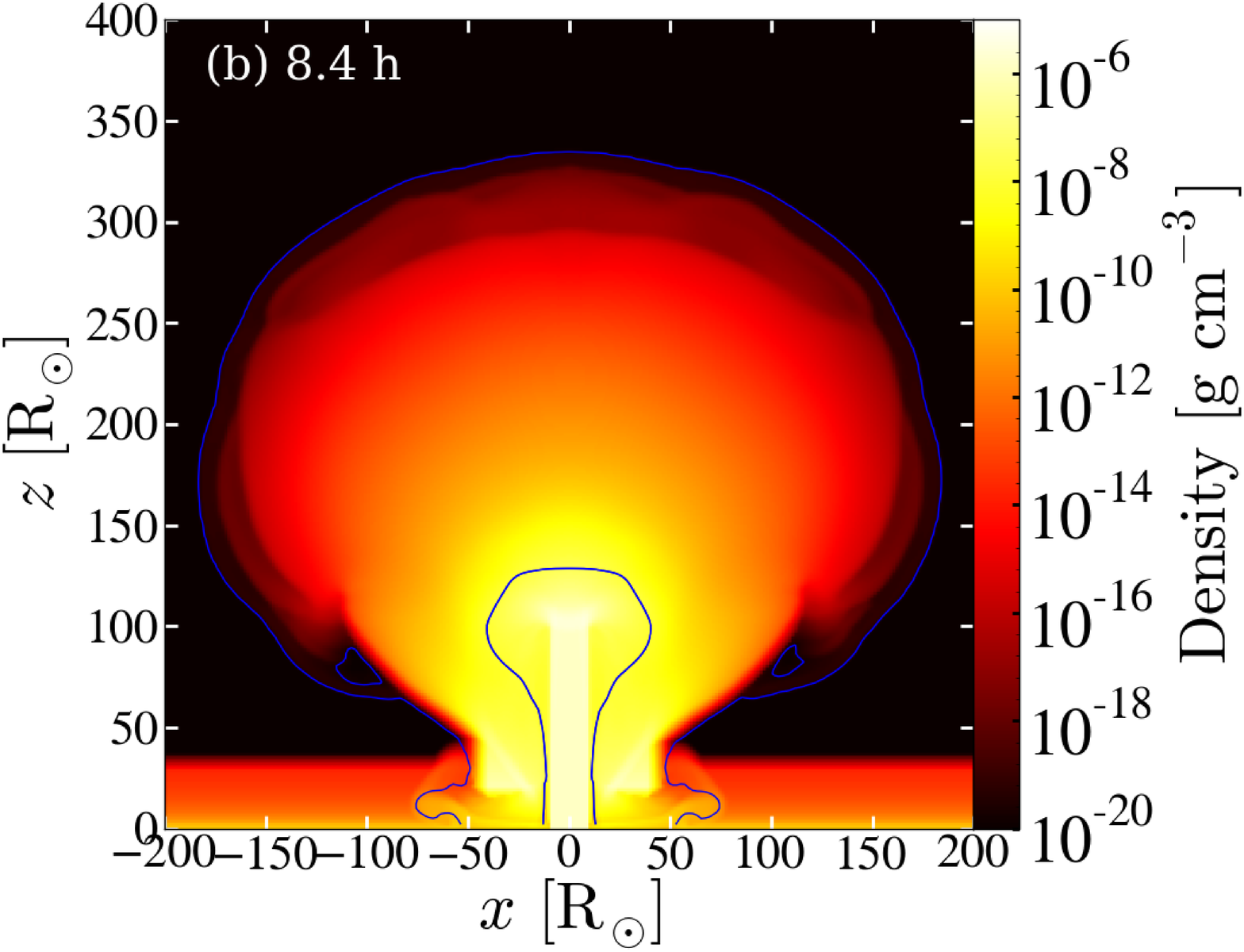}}\\
\hspace{-0.4 cm}
\subfigure{\label{subfigure:temp1}\includegraphics*[scale=0.21,clip=false,trim=0 0 0 0]{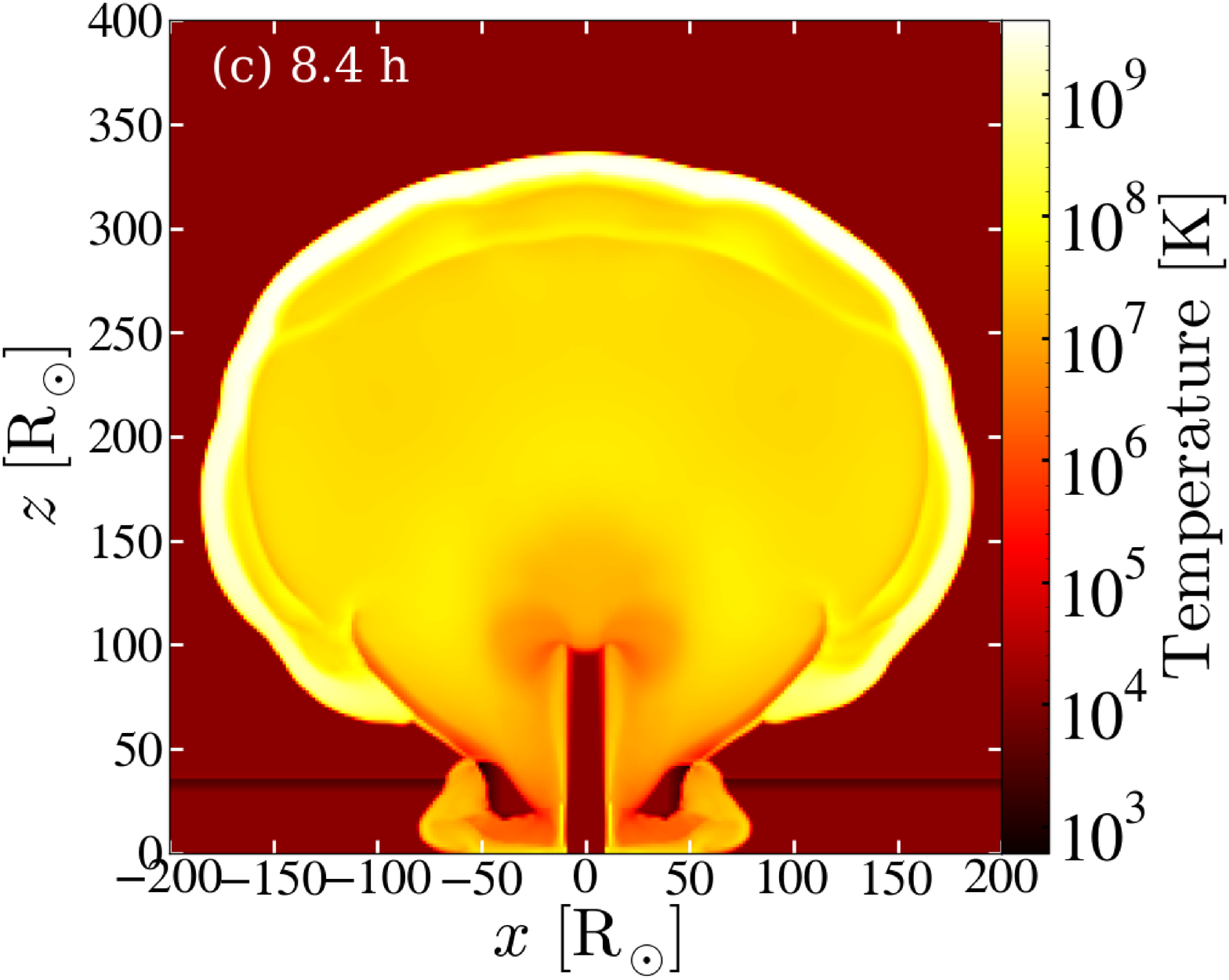}}
\hspace{0.2 cm}
\subfigure{\label{subfigure:velz1}\includegraphics*[scale=0.21,clip=true,trim=0 0 0 0]{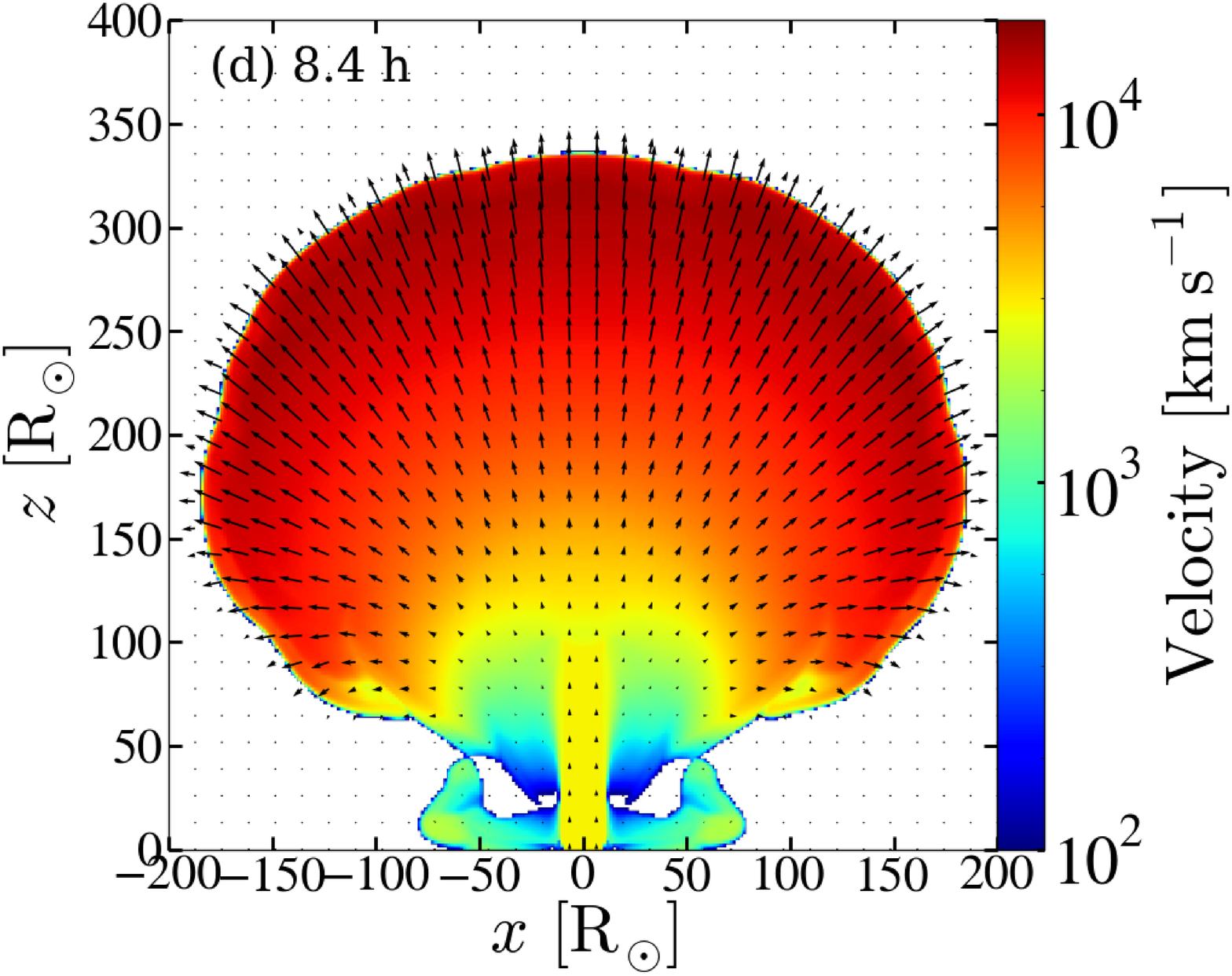}}
\caption{
Various quantities in the $y = 0$ plane for the $v_{\rm{jets}} = 3000 \km \s^{-1}$ case.
The $z = 0$ plane is located about where the outskirt of the temporary common envelope is, ${\rm several} \times 10 R_\odot$ above the equatorial plane.
Panels (a) and (b) show the density structure at two times.
The blue contour shows material originally in the shell.
(c) The temperature map.
Note that the temperature in the outer regions should be lower than presented,
as we do not include radiative cooling in our simulations.
The regions with temperature of $10^4 \K$ did not go through any interaction.
(d) The velocity map.
  }
\label{fig:results_3000}
\end{center}
\end{figure}

\begin{figure}[ht!]
\begin{center}
\hspace{-1.0 cm}
\subfigure{\label{subfigure:dens_2000}\includegraphics*[scale=0.21,clip=false,trim=0 0 0 0]{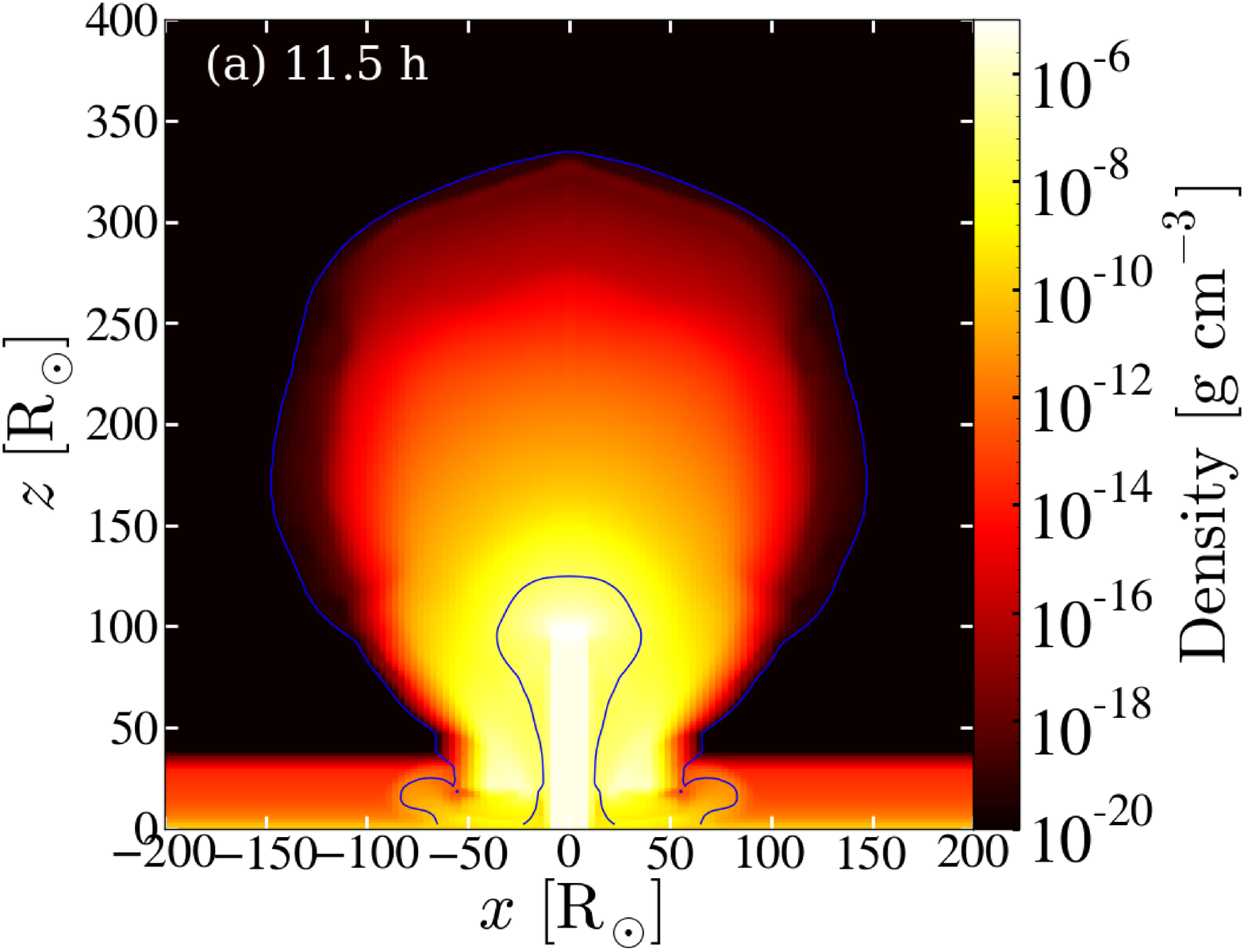}}
\subfigure{\label{subfigure:velz_2000}\includegraphics*[scale=0.21,clip=true,trim=0 0 0 0]{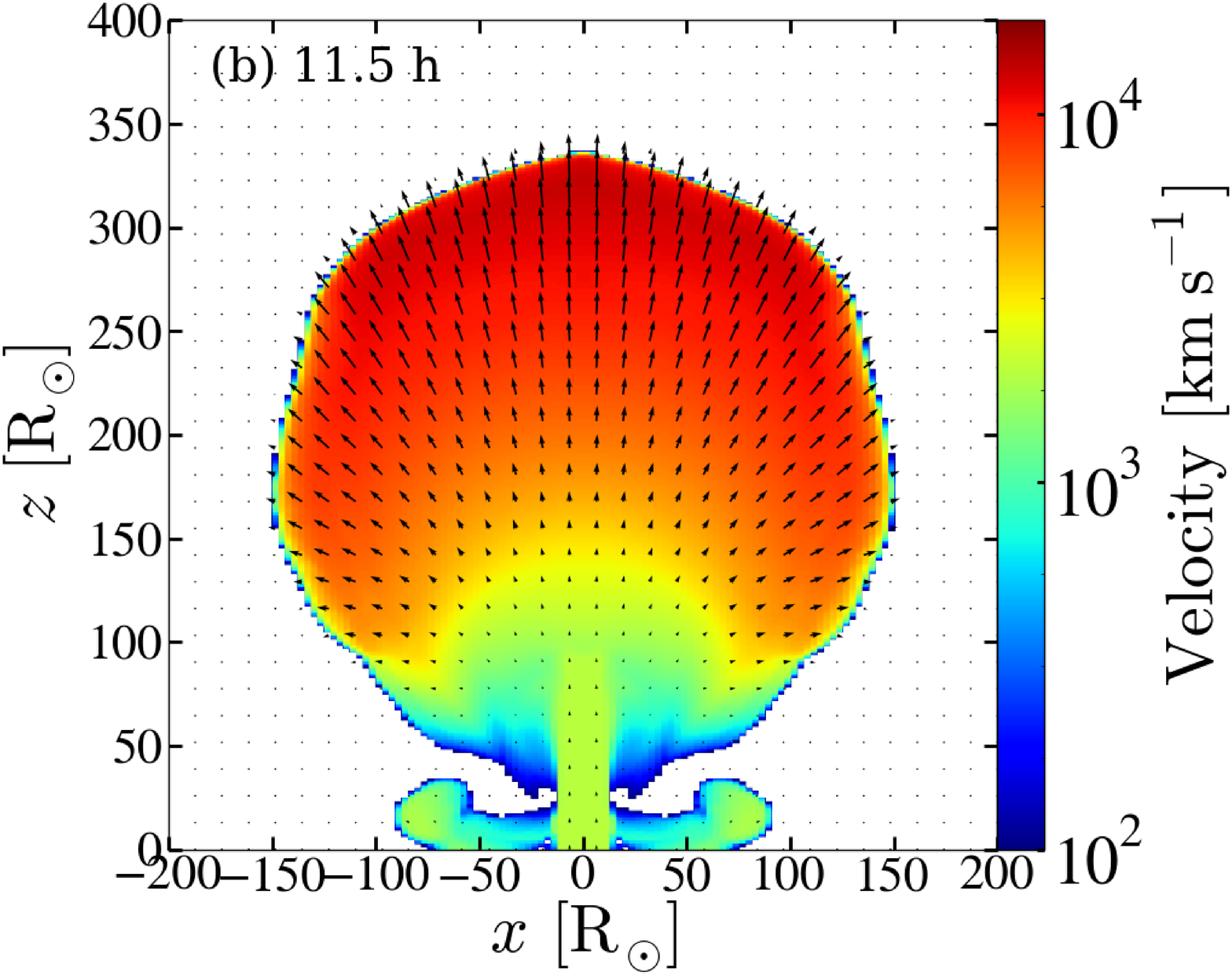}}
\caption{
Same as panels (b) and (d) in Figure \ref{fig:results_3000}, but for the $v_{\rm{jet}} = 2000 \km \s^{-1}$ case and at a later time.
}
\label{fig:results_2000}
\end{center}
\end{figure}

As the jet hits the shell a shock wave is formed.
Because of the steeply declining density the shock accelerates outward, and with it the shell material, up to velocities of $>10^4 \km \s^{-1}$.
This is demonstrated in Figure \ref{fig:results_3000} for the $v_{\rm jet} = 3000 \km \s^{-1}$ case, and in
Figure \ref{fig:results_2000} for the $v_{\rm{jet}} = 2000 \km \s^{-1}$ case.
After the jet breaks out of the shell the acceleration ceases.
However, in reality the secondary star orbits the primary star and so the jets' source moves,
allowing the jet to encounter additional shell material.
During the several hours jet-activity period the secondary will move $\sim 20 R_\odot$ near periastron \citep{Kashietal2013},
so that the total accelerated mass can be $\sim 2-3$ times larger than the mass we obtain here.

\begin{figure}[ht!]
\begin{center}
\vspace{-0.7 cm}
\subfigure{\label{subfigure:vel_hist1}\includegraphics*[scale=0.115,clip=false,trim=0 0 0 0]{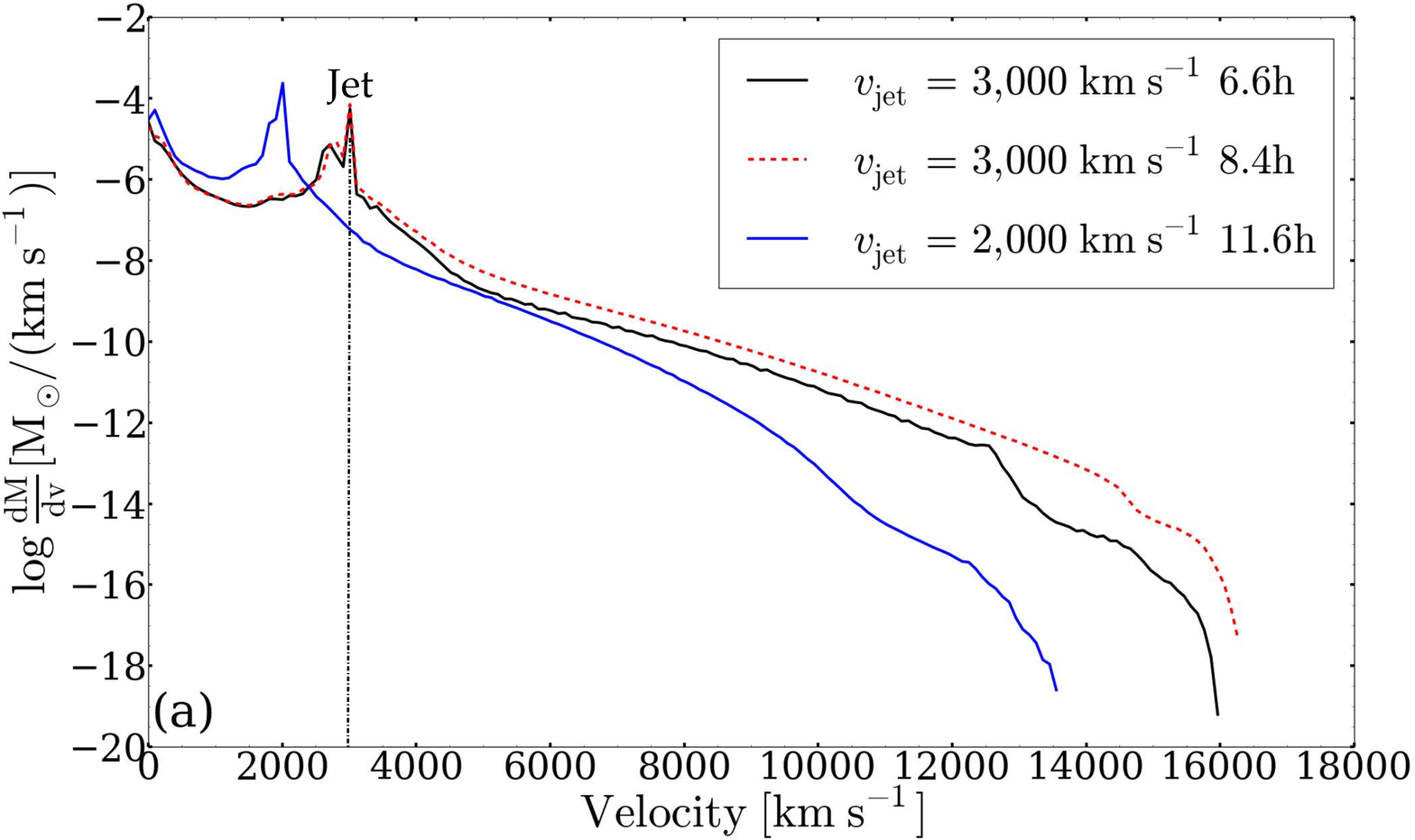}}
\vspace{-0.4 cm}
\subfigure{\label{subfigure:vel_hist2}\includegraphics*[scale=0.115,clip=false,trim=0 0 0 0]{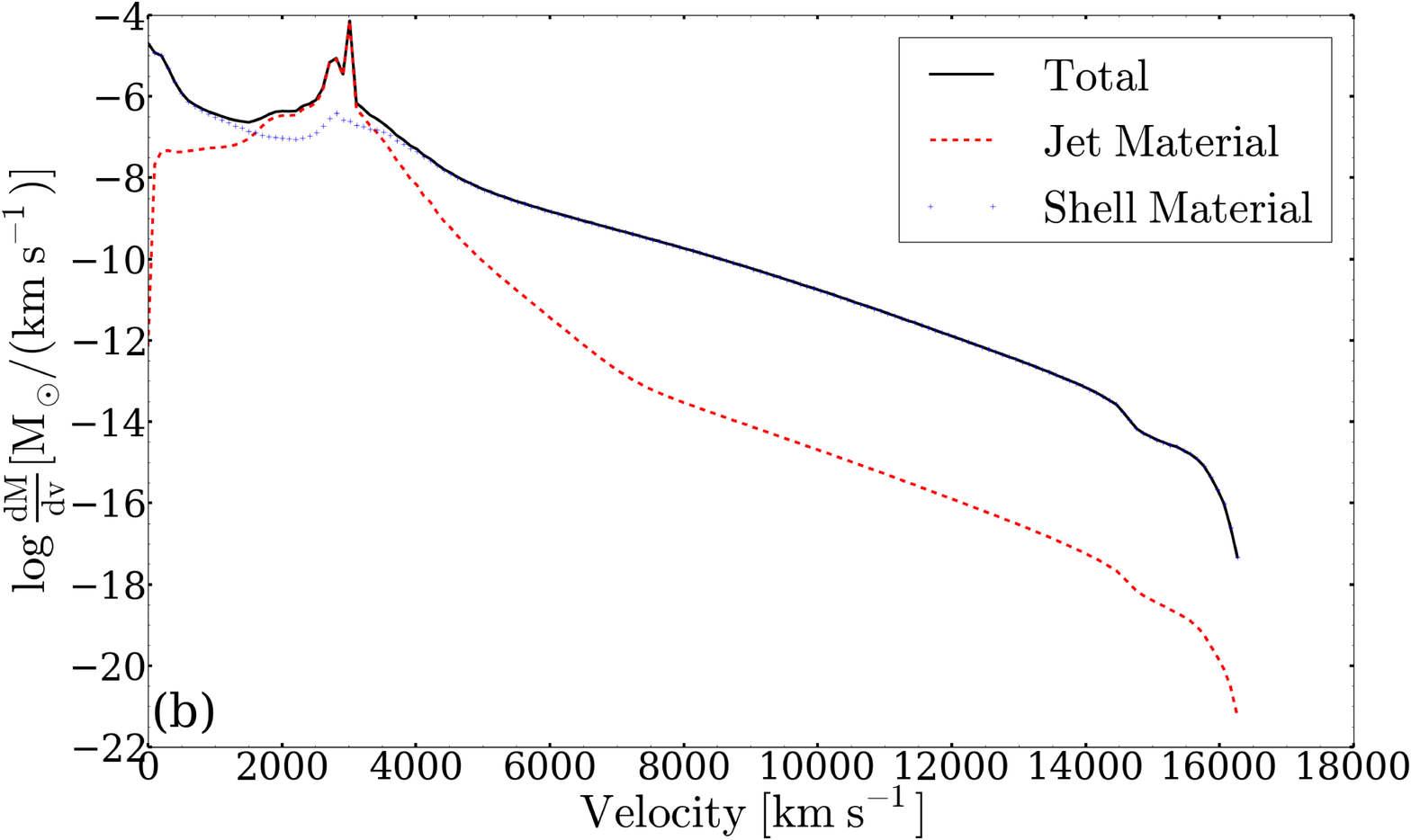}}
\vspace{-0.4 cm}
\subfigure{\label{subfigure:vel_hist3}\includegraphics*[scale=0.115,clip=false,trim=0 0 0 0]{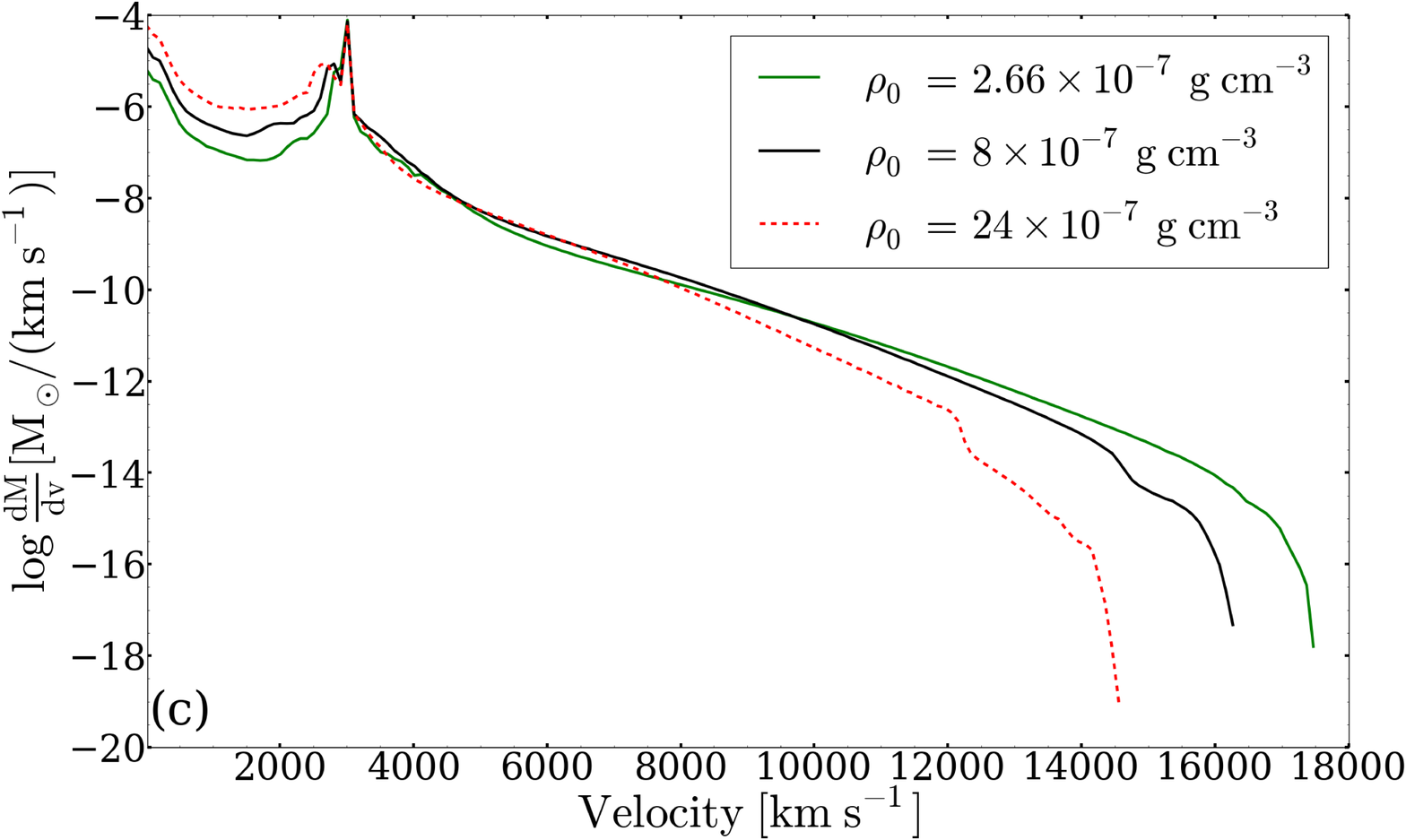}}
\vspace{-0.4 cm}
\subfigure{\label{subfigure:vel_hist4}\includegraphics*[scale=0.115,clip=false,trim=0 0 0 0]{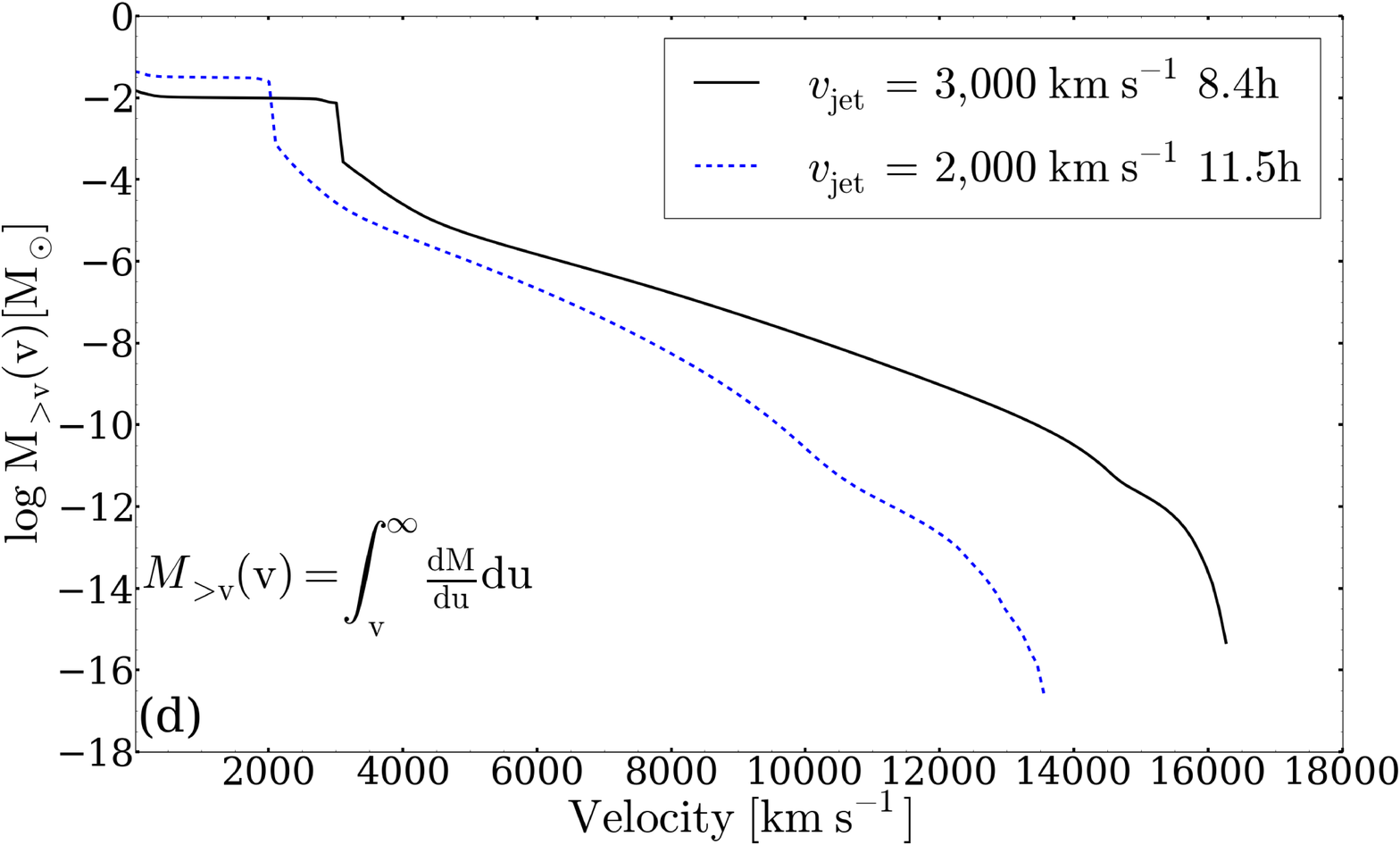}}
\caption{
Various plots for both the $3000 \km \s^{-1}$ and the $2000 \km \s^{-1} $ runs.
(a) Distribution of mass with velocity. The dashed vertical line marks $3000 \km \s^{-1}$.
(b) Distrubution of jet material and shell material constituents for the $3000 \km \s^{-1}$ run.
(c) Distribution of mass with velocity for various values of $\rho_0$.
The $\rho_0=8 \times 10^{-7} \g \cm^{-3}$ is our fiducial case whose results are presented also in Figure \ref{fig:results_3000}.
(d) Total mass above a certain velocity.
}
\label{fig:vel_hist}
\end{center}
\end{figure}

To quantify the mass distribution at different velocities we plot in Figure \ref{fig:vel_hist} the value of $dM/dv$ as function of velocity,
as well as the total mass above a certain velocity $M_{>v} (v)$.
In panel \ref{subfigure:vel_hist1},  we present $dM/dv$ for the $v_{\rm jet} = 3000 \km \s^{-1}$ case at two times as indicated,
and the same at a late time for the $v_{\rm{jet}} = 2000 \km \s^{-1}$ case.
Although reaching lower velocities than in the $v_{\rm{jet}} = 3000 \km \s^{-1}$ case,
jets of $v_{\rm{jet}} = 2000 \km \s^{-1}$ can also accelerate gas to very high velocities.
In panel \ref{subfigure:vel_hist2} we show the distribution of the original jet and shell media constituents for the $v_{\rm jet} = 3000 \km \s^{-1}$ case. The material at velocities larger than $3000 \km \s^{-1}$ is predominantly shell material.
Panel \ref{subfigure:vel_hist3} depicts the results for $v_{\rm jet} = 3000 \km \s^{-1}$ cases where the shell mass was either 3 times larger or 3 times lower than our fiducial case presented in the lower panel of Figure \ref{fig:setup}.
When the density of the shell is 3 times larger than in our fiducial case,
the inner boundary of the shell is denser than the jet.
We find that such an encounter does not lead to efficient acceleration to very high velocities.
In panel \ref{subfigure:vel_hist4} we plot $M_{>v} (v)$ for the $v_{\rm jet} = 3000 \km \s^{-1}$ case.

For the $3000 \km \s^{-1}$ case the total mass (two jets) of gas having a velocity of $ > 6000 \km \s^{-1}$ is $\sim {\rm few} \times 10^{-6} M_\odot$.
The photosphere during the pre-2012b outbursts is at $\sim {\rm few} \times 10^{14} \cm$ \citep{Marguttietal2013},
so that the hydrogen column density and density are $\sim 10^{21} \cm^{-2}$ and $\sim 10^7 \cm^{-3}$, respectively.
Is this gas sufficient to account for the absorption observed by \cite{Pastorello2012} in the 2011 outburst of SN~2009ip?
Interestingly, these values are very similar to those of the absorbing gas in the broad absorption lines (BAL) quasar LBQS~1206+1052,
where clear Balmer absorption lines are detected \citep{JiWang2012}.
Further study that includes calculation of the ionization stages of various elements is required and planned for the future,
but it seems that the amount of very high-velocity gas
obtained in our simulations can in principle account for the observations of the 2011 outburst of SN~2009ip.

We briefly comment on two of our assumptions,
that the secondary wind plays no role in the interaction, and that the radiative cooling is small.
We are confident that the wind from the secondary star plays no role during the unstable phases of the primary star,
when the primary stellar envelope expands and its mass loss rate is high.
The accretion onto the secondary star is triggered by dense clumps formed in the post-shock region of the LBV dense envelope or wind,
as was calculated and demonstrated for the accretion onto the companion of $\eta$ Carinae near periastron passages;
neither the wind nor the radiation pressure of the accreting secondary star can prevent accretion
\citep{Soker2005, KashiSoker2009, Akashietal2013}.
In $\eta$ Carinae the LBV wind density, as used in the above calculations,
is much lower than the wind density in the bloated envelope of the LBV primary in our model of SN2009ip,
and the secondary wind and radiation pressure play an even smaller role than in present $\eta$ Carinae.
As well, as in the case of the winds interaction in $\eta$ Carinae \citep{Parkinetal2011, Akashietal2013},
the wind from the LBV star bends the wind of the companion, being an O star or a WR star,
such that the wind from the secondary star does not influence at all the interaction region near the LBV star that is studied here.
 
\begin{figure}[ht!]
\begin{center}

\includegraphics*[scale=0.20,clip=false,trim=0 0 0 0]{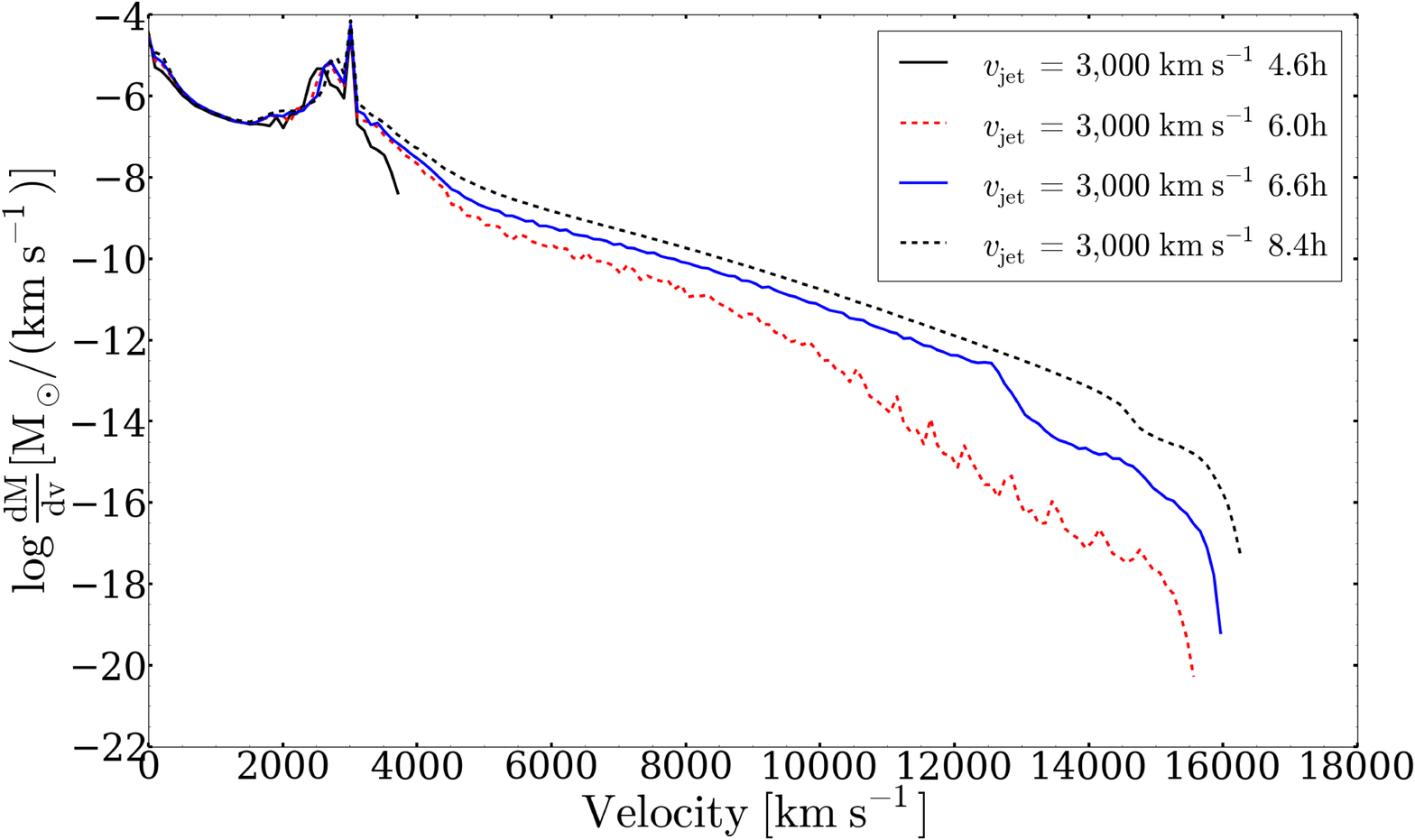}
\caption{
Distribution of mass with velocity at different times.
}
\label{fig:time_evolution}
\end{center}
\end{figure}

To explore the role of radiative cooling we compare the photon diffusion time with expansion time.
In Figure \ref{fig:time_evolution} we show the evolution of the value of $dM/dv$ as function of velocity.
The main acceleration occurs within $\sim 1~$h around $t \sim 5~$h. 
 
We take the results presented in Figure \ref{fig:results_3000} at time $6.6~$hours.
At point $(z,x)=(125 R_\odot,0)$ 
the velocity is $\sim 5,600 \km \s^{-1}$, and the density is $\sim 10^{-9.5} \g \cm^{-3}$.
The distance from this region outward is $\sim 65 R_\odot$, and for electron scattering the optical depth is $\tau \sim 500$.
The photon diffusion time out of this point is $\sim 20~$ hours, much longer than the diffusion time. 
At points $(z,x)=(140 R_\odot, 0)$ and $(z,x)=(144 R_\odot, 0)$ the velocities are $8,000 \km \s^{-1}$ and $9,000 \km \s^{-1}$,
and the diffusion times are $\sim 0.7~$h and $\sim 0.2~$h, respectively. 
These simple calculations show that radiative cooling is important and must be considered in future calculations.
For the parameters used here radiative cooling proceeds much faster than acceleration for $v \ga 8000 \km \s^{-1}$.
For gas at these velocities we overestimate the mass, but our results hold for lower velocities.
We expect that a more compact interaction region will give higher velocities.
This can be the case for narrower jets from a WR secondary star that interact closer to the launching region,
as the flow time is shorter and the diffusion time is longer. 
It is possible that these results hint to a WR secondary star if SN~2009ip is indeed a binary star. 

\section{SUMMARY}
\label{sec:summary}
There is an ongoing debate regarding the nature of the SN impostor SN~2009ip and
the several outbursts that followed its 2009 eruption.
Most controversial is the nature of the  2012a+2012b outbursts \citep{Martinetal2013},
with three main types of scenarios:
($i$) Single terminal explosion of the star (either in the 2012a or 2012b outburst).
Namely, a SN explosion which can be followed by interaction with the CSM \citep{Smithetal2013, Mauerhan2013}
or by a quark-nova \citep{Ouyedetal2013}.
($ii$) Non-terminal single star eruption \citep{Pastorello2012} with one outburst in 2012a followed by shell collision for the 2012b outburst,
or two outbursts in 2012a and 2012b.
($iii$) Binary interaction \citep{SokerKashi2013, Kashietal2013}.

In the binary model the LBV primary enters an unstable phase that triggers binary interaction near periastron passages
\citep{SokerKashi2013, Kashietal2013}.
The main energy source is the launching of two opposite jets by the secondary star as it accretes mass from the primary envelope.
As the jets are launched at $v_{\rm jet} \simeq 1-2 v_{\rm esc}$, where $v_{\rm esc}$ is the escape velocity from the secondary star,
one of the challenges for the binary model is to account for gas moving at $\ga 10,000 \km \s^{-1}$
as observed in the 2011 outburst \citep{Pastorello2012} and in the 2012a outburst \citep{Mauerhan2013}.
In the present study we addressed this challenge. 

Our flow structure that is depicted schematically in Figure \ref{fig:setup} has two basic assumptions.
($i$) A binary companion launches jets in the outskirts of the primary envelope.
This is taken from the binary model \citep{SokerKashi2013, Kashietal2013}.
($ii$) Gas with a steep density profile resides within $\sim 1 \AU$ around the binary system.
The shell might actually be composed of clouds and segments.
The total mass in such a shell is $\sim 0.1 M_\odot$.
We expect such an extended envelope to be formed from the unstable primary phase and the strong binary interaction.
Our assumption of the existence of this shell (extended envelope), will be examined in a future work.

We numerically studied the propagation of the jets through the extended envelope.
In the present demonstrative paper we simulate only two types of jets,
with velocities of $v_{\rm jet} =3000 \km \s^{-1}$ and $v_{\rm jet} =2000 \km \s^{-1}$,
and vary only the density of the shell, not its size or profile.
The results of our simulations for initial jet velocities of $v_{\rm{jet}} = 3000 \km \s^{-1}$ and $v_{\rm{jet}} = 2000 \km \s^{-1}$
are presented in Figures \ref{fig:results_3000} and \ref{fig:results_2000}, respectively.
While the jet is well collimated, the interaction with the shell creates an outflowing gas with a large opening angle.
Although the flow remains non-spherical, i.e., has a bipolar structure when both jets are considered,
no fine tuning of the observing angle is required in order to observe the absorption lines.
After the gas expands to more than several $\AU$, the absorption lines should be detected from most directions.

We found that for the parameters used here radiative cooling will reduce the amount of gas
with velocities of $\ga 8000 \km \s^{-1}$.
Jets launched by a WR companion will be narrower and denser than the jets simulated here,
with shorter flow time and longer photon diffusion time.
This will allow acceleration to higher velocities.
We take this as a hint that in the binary model of SN~2009ip the secondary star should be a WR star \citep{Kashietal2013}.

In section \ref{sec:results} we estimated that there is sufficient gas to explain the observation of \cite{Pastorello2012}
of absorption lines at $>10^4 \km \s^{-1}$ in the 2011 outburst.
The energy and duration of the 2012a+b outbursts are $\sim 30$ times those of the 2011 outburst. 
Scaling from several hours to several days, we deduce that our flow structure can account for $\sim 10^{-4} M_\odot$ of gas with $v \ga 6000 \km \s^{-1}$ in the 2012b outburst.
It is desired to explore the parameters space to find a more efficient acceleration flow structure, e.g., taking a WR companion,
such that more mass will be accelerated to very fast velocities.
This is the task of an ongoing research.
Our present results could offer an explanation to the nature of the 2011 outburst,
and further support the claim made by \cite{SokerKashi2013} and \cite{Kashietal2013} regarding the binary nature of SN~2009ip system.

\vspace{1 cm}
We thank the anonymous referee for useful comments.

{}


\begin{thebibliography}{}

\bibitem[Akashi et al.(2013)]{Akashietal2013} Akashi, M.~S., Kashi, A., \& Soker, N.\ 2013, \na, 18, 23

\bibitem[Berger et al.(2009)]{Berger2009} Berger, E., Foley, R., \& Ivans, I.\ 2009, Astron., Tel., 2184, 1

\bibitem[Davidson \& Humphreys(1997)]{DavidsonHumphreys1997} Davidson, K., \& Humphreys, R.~M.\ 1997, \araa, 35, 1

\bibitem[Drake et al.(2012)]{Drake2012} Drake, A.~J., et al.\ 2012, Astron., Tel., 4334, 1

\bibitem[Foley et al.(2011)]{Foleyetal2011} Foley, R.~J., Berger, E., Fox, O., et al.\ 2011, \apj, 732, 32 

\bibitem[Fryxell et al.(2000)]{Fryxell2000} Fryxell B., Olson K., Ricker P., et al.,\ 2000, \apjs, 131, 273

\bibitem[Ireland(2011)]{Ireland2011} Ireland, M.~J.\ 2011, Why Galaxies Care about AGB Stars II: Shining Examples and Common Inhabitants, 
445, 83 

\bibitem[Ji et al.(2012)]{JiWang2012} Ji, T., Wang, T.-G., Zhou,  H.-Y., \& Wang, H.-Y.\ 2012, Research in Astronomy and Astrophysics, 12, 369

\bibitem[Kashi \& Soker(2009)]{KashiSoker2009} Kashi, A., \& Soker, N.\ 2009, \na, 14, 11

\bibitem[Kashi \& Soker(2010)]{KashiSoker2010} Kashi, A., \& Soker, N.\ 2010, \apj, 723, 602

\bibitem[Kashi et al.(2013)]{Kashietal2013} Kashi, A., Soker, N., \& Moskovitz, N.\ 2013, arXiv:1307.7681

\bibitem[Levesque et al.(2012)]{Levesque2012} Levesque, E.~M., Stringfellow, G.~S., Ginsburg, A.~G., Bally, J., \& Keeney, B.~A.\ 2012, arXiv:1211.4577

\bibitem[Margutti et al.(2013)]{Marguttietal2013} Margutti, R., Milisavljevic, D., Soderberg, A.~M., et al.\ 2013, arXiv:1306.0038

\bibitem[Martin et al.(2013)]{Martinetal2013} Martin, J.~C., Hambsch, F.-J., Margutti, R., et al.\ 2013, arXiv:1308.3682

\bibitem[Mauerhan et al.(2013)]{Mauerhan2013} Mauerhan, J.~C., Smith, N., Filippenko, A.~V., et al.\ 2013, \mnras, 430, 1801

\bibitem[Maza et al.(2009)]{Maza2009} Maza, J., et al.\ 2009, CBET, 1928, 1

\bibitem[Ohnaka(2013)]{Ohnaka2013} Ohnaka, K.\ 2013, \aap, 553, A3 

\bibitem[Ouyed et al.(2013)]{Ouyedetal2013} Ouyed, R., Koning, N., \& Leahy, D.\ 2013, arXiv:1308.3927

\bibitem[Parkin et al.(2011)]{Parkinetal2011} Parkin, E.~R., Pittard, J.~M., Corcoran, M.~F., \& Hamaguchi, K.\ 2011, \apj, 726, 105

\bibitem[Passy et al.(2012)]{Passyetal2012} Passy, J.-C., De Marco, O., Fryer, C.~L., Herwig, F., Diehl, S., Oishi, J.~S., Mac Low, M.-M.,
        Bryan, G.~L., \& Rockefeller, G. 2012, \apj, 744, 52

\bibitem[Pastorello et al.(2012)]{Pastorello2012} Pastorello, A., Cappellaro, E., Inserra, C., et al.\ 2013, \apj, 767, 1

\bibitem[Pittard \& Corcoran(2002)]{PittardCorcoran2002} Pittard, J.~M., \& Corcoran, M.~F.\ 2002, \aap, 383, 636

\bibitem[Prieto et al.(2013)]{Prieto2013} Prieto, J.~L., Brimacombe, J., Drake, A.~J., \& Howerton, S.\ 2013, \apjl, 763, L27

\bibitem[Ricker \& Taam(2012)]{RickerTaam2012} Ricker, P.~M., \& Taam, R.~E.\ 2012, \apj, 746, 74

\bibitem[Smith(2008)]{Smith2008} Smith, N.\ 2008, \nat, 455, 201

\bibitem[Smith et al.(2011)]{Smithetal2011} Smith, N., Li, W., Silverman, J.~M., Ganeshalingam, M., \& Filippenko, A.~V.\ 2011, \mnras, 415, 773

\bibitem[Smith et al.(2013)]{Smithetal2013} Smith, N., Mauerhan, J.,\& Prieto, J.\ 2013, arXiv:1308.0112

\bibitem[Smith et al.(2010)]{Smithetal2010} Smith, N., Miller, A., Li, W., et al.\ 2010, \aj, 139, 1451

\bibitem[Soker(2005)]{Soker2005} Soker, N.\ 2005, \apj, 635, 540

\bibitem[Soker(2008)]{Soker2008} Soker, N.\ 2008, \na, 13, 491

\bibitem[Soker \& Kashi(2013)]{SokerKashi2013} Soker, N., \& Kashi, A.\ 2013, \apjl, 764, L6

\bibitem[Weis et al.(2001)]{Weisetal2001} Weis, K., Duschl, W.~J., \& Bomans, D.~J.\ 2001, \aap, 367, 566

\end{thebibliography}
\end{document}